# COMPUTER-AIDED PLANNING FOR ZYGOMATIC BONE RECONSTRUCTION IN MAXILLOFACIAL TRAUMATOLOGY

S. MAUBLEU (1), CH. MARECAUX (1,2), M. CHABANAS (1), Y.PAYAN (1), F. BOUTAULT (2)

(1) TIMC-IMAG Laboratory, Institut de l'Ingénierie et de l'Information de Santé, Grenoble 38700 La Tronche, France
(2) Department of Maxillofacial and Facial Plastic Surgery, CHU Purpan, place Baylac 31059 Toulouse, France

A common maxillofacial trauma involves the fracture and dislocation of the zygomatic bone, with severe morphological dysmorphosys. It can be associated with a "blow-out fracture" of the orbital floor, leading to a limitation of the upper eye movement, ocular injury or enophthalmia. The surgical correction of these fractures is undertaken at approximately 3-5 days after injury to allow swelling to subside. While the aim of the procedure is to ensure the post-operative aesthetic of the patient to be as close as possible to its pre-trauma state, it is quite difficult to achieve due to the small operating field, the lack of anatomical reference location, and the important swelling.

One of the topic of the computer-aided maxillofacial projects developed in our group [3,4] is then to address the correction of these fractures of the Zygomatic bone, which includes the definition of an optimal surgical planning and the development of a specific navigation system for intra-operative guiding.

This paper focuses on the first step, namely defining an optimal surgical planning for the reconstruction of the fractured midface side.

## METHOD

The aim of the surgical procedure is to reconstruct the facial skeleton as close as possible to how it was before the fracture. Since no pre-traumatic data are available, the only reference to the original patient morphology is the healthy side of the face [6]. Although it is not completely true, the facial skeleton of normal subjects can be reasonably assumed as symmetric. Therefore, an optimal surgical goal is to reposition the fractured zygomatic bone symmetrically with respect to the unaffected side of the patient. A planning protocol is proposed, divided in four steps: three-dimensional reconstruction from CT imaging, bone fragment segmentation, mirroring of the facial skeleton, and fragment registration.

## 1. 3D reconstruction from CT imaging

The computer-aided protocol rely on a CT scan of the midface. 3D models of the skin surface and the facial skeleton are first reconstructed using a marching cubes algorithm [2]. Thresholds for isosurface generation are automatically set with reference Housnfield units, although they can be manually set to improve the segmentation of the region of interest.

## 2. Fragment segmentation

Different methods, like thresholding, mathematical morphology, and region growing algorithms, were investigated to automatically segment the fractured bone in the CT data. However, the zygomatic fragment is generally still connected with the rest of the skeleton, thus very difficult to isolate. We have finally chosen to manually extract the fragment using a sphere define by four points picked on the 3D model (fig. 1), which rather correspond to the fracture points on the maxillary, orbital, fronto-sphenoidal and zygomatic processes.

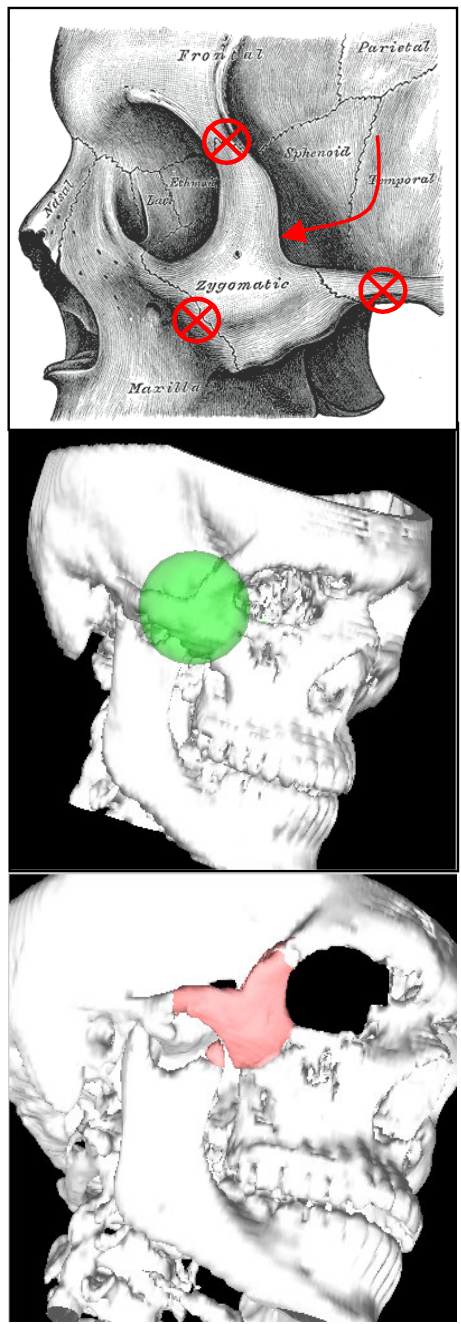

*Figure 1: segmentation of the fractured zygomatic bone*

This sphere always contains most of the zygomatic bone. To ensure no additional structures are segmented, e.g. a bone splinter or a part of coronoid process (fig. 1), only the greatest object within the sphere, the malar fragment, is kept. The fracture boundaries may not be present, but they are generally useless since they are rarely sharply defined.

The segmented zygomatic bone is the object that will have to be repositioned during the surgery. For the next step of the planning, the skull mirroring, it is removed from the 3D model to only keep the non-affected structures surrounding the fracture.

## 3. Skull Mirroring

The target position for the fractured bone fragment is defined by mirroring the healthy side of the skeleton around the mid-sagittal plane.

In axial CT slices, this plane is in theory orthogonal to the slices and thus straightforward to compute. However, the patient's head is always tilted in the device during the scanning process, which makes the images asymmetric and quite difficult to analyse. The actual mid-sagittal plane must therefore be computed out of the patient anatomy. The surgeon can browse through the images (the native axial slices, plus the computed sagittal and coronal views) and the 3D models to manually identify several anatomical landmarks that belong to the mid-sagittal plane. These landmarks are the foramen caecum, the posterior extremity of the sphenoid crest and the middle point between both apophysis clinoid. They can be considered as part of the anatomical mid-sagittal plane, and are not affected by fractures of the facial skeleton. Moreover, they are reliable, precise and quite easy to locate on CT data. A first estimation of the plane is then computed out of these three points.

Using this initial sagittal plane, each vertex of the healthy side 3D model is mirrored to generate a pseudo-symmetrical skull (fig.2, center). This operation is done only once, and requires up to thirty seconds for 185 slices.

To improve the accuracy of the mirroring, a registration algorithm [7] is used to match the mirrored healthy side with the bone structures surrounding the fractured zygomatic fragment. These regions of interest are the orbital margin, the zygomatic process and the nasal area if it has not been dislocated with the fracture. They are manually defined by the surgeon, who just has to click three or four points on the 3D model to determine their centre. A rigid registration is then performed, which enable to overcome the errors inherent to the limited accuracy of the manual plane definition. Thus, an elastic registration enables to account for the natural asymmetries of the facial skeleton, and ensures the continuity between the target and the remaining bone structures, in the zygomatic and orbital areas (fig.2, right). These two registration steps are automatic, fast, and quite robust since the initial position given by the initial mirroring process is excellent.

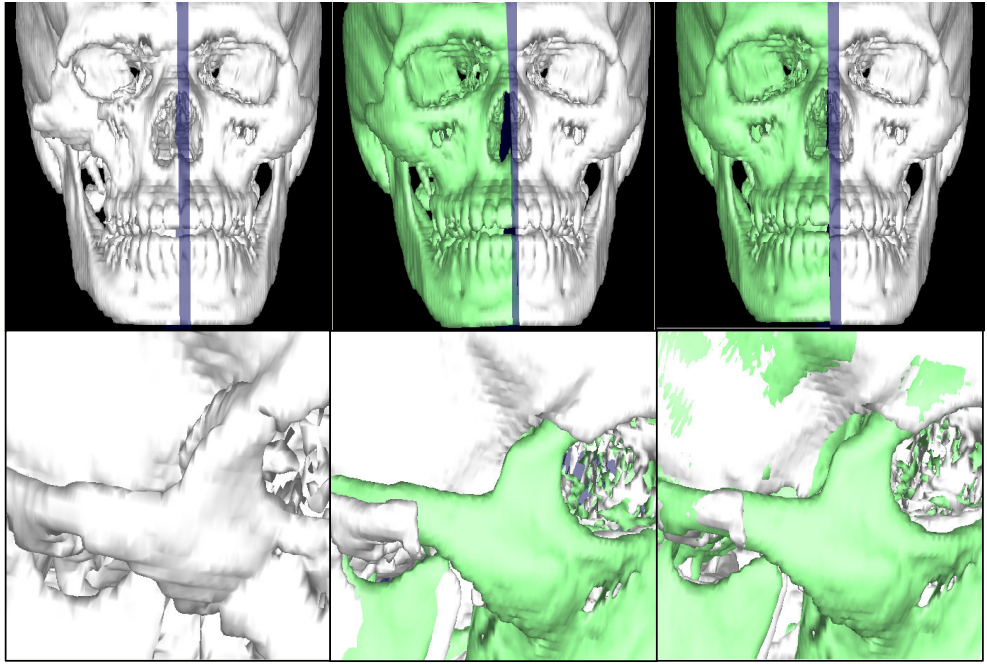

*Figure 2 : From left to right, Real patient skull, Pseudo-symmetrical skull and registered pseudo-symmetrical skull. White skull is the real skull, and Grey skull is the mirrored target. The lower panel shows that the continuity between the target and the surrounding bone structures is ensured, in the zygomatic and orbital areas.*

Thanks to this skull mirroring process, an estimation of the zygomatic bone before the fracture has been obtained, which will be used to guide the surgeon to replace the fractured bone during the surgery.

### *4. Fragment registration*

To evaluate the fracture displacement and the correction to apply, a rigid registration can be performed between the segmented zygomatic fragment and its target position (fig. 3). According to the fragment dislocation, a manual pre-registration may be needed.

## RESULTS

The planning process has first been carried out on four patients suffering from fractures of the zygomatic bone and/or the orbital floor. Despite some manual interactions are required, the overall planning time never exceeded 10 minutes, which is therefore compatible with a use in clinical routine.

Beside the feasibility and the user-friendly character of the application, the accuracy of the mirrored target has also been evaluated on CT scans of seven healthy subjects that do not suffer mid-facial fractures. For each subject, the zygomatic bone can be removed on one side of the face, to simulate patient data after the fragment segmentation step. After the mirroring procedure is performed, the distance between the surface of the computed target and the actual zygomatic bone is computed.

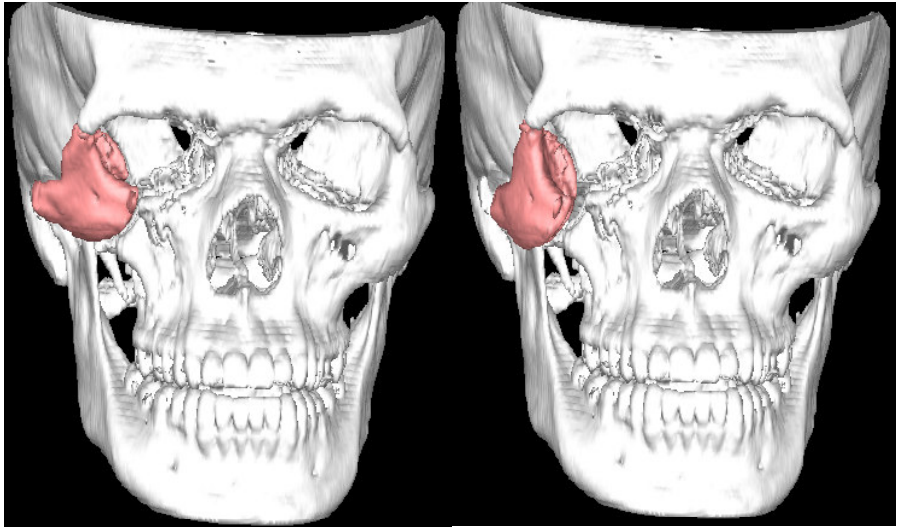

*Figure 3: initial position of the zygomatic fragment (left) and final planned position after registration to the mirrored target surface (right).*

This validation protocol was applied on seven healthy subjects. Considering either right or left mid-face side is healthy, this actually provided fourteen "patients". Five of the subjects have an equilibrated skull, while two of them suffer from a dysmorphosys or a non-symmetrical growth of the skeleton, which is a good test to evaluate the ability of the method to account for the natural asymmetries of the skull. The computed mean errors (tab. 1) are 1.06 mm in mean, with a maximum a 2.23 mm. The maximal errors never exceed 3.45 mm. There is no significant difference between the patients suffering from a facial dysmorphosys and the others.

These results show that the assumption of the facial skeleton symmetry seems a reasonable base for the planning, even for non-equilibrated patients. Moreover, the further local registration provides a very good final target position for the zygomatic fragment repositioning.

## CONCLUSION

An optimal planning procedure has been proposed to define the target position of the zygomatic bone following a fracture of the mid-face skeleton. The protocol has been successfully tested on healthy subjects, and ensures the global symmetry of the face could be obtained after the reconstructive surgery.

| Patient (fractured side) | Number of slice | Mean Distance between Original fragments and registered fragment (mm) | Max Distance between Original fragments and registered fragment (mm) | Time (Min) |
|---|---|---|---|---|
| ARM_020212    (right) | 152 | 1.06 | 2.05 | 8 |
| ARM_020212    (left) | 152 | 1.31 | 1.93 | 9 |
| TIJP_020218    (right) | 139 | 1.65 | 3.03 | 8 |
| TIJP_020218    (left) | 139 | 1.13 | 2.42 | 6 |
| GAR_020221    (right) | 143 | 1.88 | 2.65 | 5 |
| GAR_020221    (left) | 143 | 0.40 | 1.76 | 6 |
| KRM_020219    (right) | 147 | 1.31 | 3.45 | 6 |
| KRM_020219    (left) | 147 | 0.48 | 1.04 | 6 |
| VAN_020222    (right) | 161 | 1.23 | 2.55 | 7 |
| VAN_020222    (left) | 161 | 0.2 | 0.35 | 7 |
| AM_001122    (right) | 129 | 0.56 | 2.56 | 7 |
| AM_001122    (left) | 129 | 2.23 | 3.25 | 6 |
| CA_010724    (right) | 163 | 0.95 | 1.54 | 9 |
| CA_010724    (left) | 163 | 0.58 | 1.02 | 7 |
| Mean Values |  | 1.06 | 2.11 | 7 |

*Table 1 : distances computed between the target surface and the actual zygomatic bone surface, on seven subjects that do not suffer mid-facial fractures. The last two subjects present a natural dysmorphosys or a non-symmetrical growth of the skeleton.*

Now that the planning procedure is available, the next step of this project will be to develop an intra-operative guiding system to help the surgeon to follow the planning [1]. This procedure will mainly rely on the intra-operative registration of the zygomatic bone fragment, and the design of specific surgical ancillaries for cranio-maxillofacial surgery.

# ACKNOWLEDGEMENT

This project is supported as a "Projet Hospitalier de Recherche Clinique", region Midi-Pyrénées, Toulouse, France.

## *BIBLIOGRAPHY*

**[1] HAßFELD S., MÜHLING J., ZÖLLER J.:**
*Intraoperative navigation in oral and maxillofacial surgery. Int J Oral Maxillofac Surg 1995, 24, S. 111-119.*

**[2] LORENSEN W.E., CLINE H.E.**
*Marching Cubes: a high resolution 3D surface construction algorithm. Computer Graphics 1987, 21:163-169.*

**[3] MARECAUX CH., CHABANAS M., LUBOZ V., PEDRONO A., CHOULY F., SWIDER P., PAYAN Y., BOUTAULT F.**
*Maxillofacial computer aided surgery: a 5 years experience and future. SURGETICA, 2002, pp. 185-190.*

**[4] MARECAUX CH., CHABANAS M., PAYAN Y., BOUTAULT F.**
*Chirurgie Assistée par Ordinateur et Chirurgie Maxillo Faciale : Principes et rappels techniques. Revue de Stomatologie et Chirurgie Maxilofaciale, 2005 (in French).*

**[5] SCHRAMM A., GELLRICH NC., GUTWALD R., THOMA L., SCHMELZEISEN R.:**
*Reconstructive computer assisted surgery of deformities by mirroring CT data sets. Med Biol Eng Comp, 1999, 37, S. 974-975.*

**[6] SCHRAMM A., GELLRICH N.C., GUTWALD R., SCHIPPER J., BLOSS H., HUSTED H., SCHMELZEISEN R., OTTEN J.E.**
*Indications for computer-assisted treatment of cranio-maxillofacial tumors. J. Computer Aided Surgery, 5(5):343-352,2000.*

**[7] SZELISKI, R., LAVALLEE, S.**
*Matching 3-D anatomical surfaces with non-rigid deformations using octree-splines. Int. J. of Computer Vision, 1996, 18(2) :171-186.*